\documentstyle[amssymb,prd,aps,tighten,epsfig]{revtex}
\newcommand{\be}{\begin{equation}}
\newcommand{\ee}{\end{equation}}
\newcommand{\ben}{\begin{eqnarray}
\displaystyle}
\newcommand{\een}{\end{eqnarray}}

\begin{document}
\draft
\title{Quantized massive scalar fields in the spacetime of a charged dilatonic
black hole}
\author{Jerzy Matyjasek}
\address{Institute of Physics, \\
Maria Curie-Sk\l odowska University,\\
pl. Marii Curie - Sk\l odowskiej 1, 20-031 Lublin, Poland\\
matyjase@tytan.umcs.lublin.pl\\
jurek@kft.umcs.lublin.pl}
\maketitle

\begin{abstract}
Employing the approximate effective action constructed from the coincidence
limit of the Hadamard-Minakshisundaram-DeWitt (HaMiDeW) coefficient $a_{3},$
the renormalized stress-energy tensor of the quantized massive scalar field
in the spacetime of a static and electrically charged dilatonic black hole
is calculated. Special attention is paid to the minimally and conformally
coupled fields propagating in geometries with $a=1,$ and to the power
expansion of the general stress-energy tensor for small values of charge. A
compact expression for the trace of the stress-energy tensor is presented.
\end{abstract}


\preprint{} \baselineskip=18pt

\section{introduction}

According to our present understanding the physical content of quantum field
theory formulated in a spacetime describing black hole is contained in the
renormalized stress-energy tensor, $\langle T^{ab}\rangle ,$ evaluated in a
physically motivated state \cite{pcwd}. And although interesting in its own,
the stress-energy tensor plays a crucial role in various applications, most
important of which is the problem of back reaction on the metric. Indeed,
treating the stress-energy tensor as a source term of the semi-classical
Einstein fields equations, one may, in principle, investigate the evolution
of the system unless the quantum gravity effects become dominant.
Unfortunately, this programme is hard to execute as the semi-classical field
equations comprise rather complicated set of nonlinear partial differential
equations, and, moreover, it requires knowledge of functional dependence of $%
\langle T^{ab}\rangle $ on a wide class of metrics. It is natural therefore
that in order to answer -- at least partially -- this question, one should
refer either to approximations or to numerical methods.

It seems that for the massive fields in a large mass limit, i.e., when the
Compton length, $l_{C},$ is much smaller than the characteristic radius of
curvature, ${\Bbb L},$ (where the latter means any characteristic length
scale of the spacetime), the approximation based on the asymptotic
Schwinger-DeWitt expansion is of the required generality~\cite
{Julian,Bryce1,Bryce2}. Since the nonlocal contribution to the effective
action could be neglected it is expected that the method yields reasonable
results provided the gravitational field is weak and its temporal changes
remain small. Despite of the above restrictions there are still a wide class
of geometries in which the approximation could be successfully applied. 

For a neutral massive scalar field with an arbitrary curvature coupling
satisfying 
\begin{equation}
\left( \Box \,-\,\xi R\,-\,m^{2}\right) \phi \,=\,0,  \label{wave}
\end{equation}
where $\xi $ is the coupling constant and $m$ is the mass of the field, the
approximate renormalized effective action, $W_{R},$ may be expanded in
powers of $m^{-2}$ \cite{fz1,fz2,fz3}. The n-th term of the expansion
involves coincidence limit of the `HaMiDeW' coefficient $\left[ a_{n}\right] 
$ \cite{Gibbons1} constructed solely from the curvature tensor its covariant
derivatives up to $2n-2$ order and contractions \cite
{Bryce1,sakai,gil1,gil2,AvraPhD,Avraros,Avranucl,Amsterdamski}. As the
complexity of the `HaMiDeW' coefficients rapidly grows with increasing $n$
their practical use is limited to $n=3,$ perhaps $n=4$. Moreover, it should
be emphasised that the Schwinger-DeWitt expansion is asymptotic and adding more
terms does not necessarily improve the approximation. Here we shall confine
ourselves to the simplest yet calculationally involved case $n=3,$ in which
the approximate effective action could be written as
\begin{equation}
W_{R}=\frac{1}{32\pi ^{2}}\int d^{4}x\sqrt{g}\frac{1}{m^{2}}\left[ a_{3}%
\right].  \label{weff}
\end{equation}

Having at one's disposal the approximation of the renormalized effective
action, the stress-energy tensor could be evaluated by means of the standard
formula 
\begin{equation}
\frac{2}{\sqrt{g}}\frac{\delta }{\delta g_{ab}}W_{R}=\langle T^{ab}\rangle .
\end{equation}
Since the coefficient $\left[ a_{3}\right] $ is rather complicated so is the
stress-energy tensor and the question arose of a practical applicability of
the thus obtained results. Fortunately it could be used in a number of
physically interesting and important cases. The method has been employed by
Frolov and Zel'nikov in a series of papers \cite{fz1,fz2,fz3} devoted to
construction of $\langle T^{ab}\rangle $ of the massive scalar, spinor and
vector fields in vacuum type-D spacetimes and generalized recently to
arbitrary geometries in \cite{kocio1,kocio2}. General formulas describing $%
\langle T^{ab}\rangle$ consist of over 100 local terms.

The effective action technique that we employ in this paper requires the 
metric of the spacetime to be positively defined. Hence, to obtain the physical 
stress-energy tensor one has to analytically continue at the final 
stage of calculations its Euclidean counterpart.

An alternative approach based on the WKB approximation of the solutions to
the radial equation and summation of the mode functions has been developed
by Anderson, Hiscock, and Samuel \cite{AHS}, who, among other things,
succeeded in construction of the general form of the stress-energy tensor of
the scalar field in a large mass limit in a static and spherically-symmetric
geometry. Both approaches give, as expected, identical results and the
detailed numerical analyses carried out by this authors show that for $%
mM\gtrsim 2$ ($M$ is the black hole mass) the accuracy of the
Schwinger-DeWitt approximation in the Reissner-Nordstr\"{o}m geometry is
quite good (1\% or better)~\cite{paul2}. The Schwinger-DeWitt method has
been employed in various contexts in \cite
{kocio1,kocio2,AHS,paul2,wormhole,inside,kocio3,kocio4,Kof1,Kof2}. The case
of the massive spinor field is currently actively investigated \cite
{paulhalf}.

In this article we shall study the stress-energy tensor of the quantized
massive scalar field with an arbitrary curvature coupling in a background of
the charged dilatonic black holes which is the solutions of the coupled
system of Einstein-Maxwell-dilaton equations. A complete set of this equations
may be easily derived from the action: 
\begin{equation}
S=\int d^{4}x\,\sqrt{-g}\left[ R-2\left( \nabla \phi \right) ^{2}-e^{-2a\phi
}F^{2}\right] ,  \label{string_action}
\end{equation}
where $\phi $ is the massless dilatonic field, $F$ is the strength of the
Maxwell field ($F_{ab}=2\nabla _{\lbrack a}A_{b]}$) and $a$ is the coupling
constant. For each value of the parameter $a$ there exists a black hole
solution depending on the electric charge and the mass~\cite
{gibbons2,horowitz}. The choice $a=1$ corresponds to low energy limit of the
string effective action, $a=\sqrt{3}$ to the four dimensional effective
model reduced from the Kaluza-Klein theory in five dimensions, and the
Einstein-Maxwell system is obtained with $a=0.$ Here we ignore the higher
curvature contribution to $S$ \cite{Myers1,Myers2}, as for example the
Gauss-Bonnet term.

Various properties of charged dilatonic black holes have been examined in a
numerous papers. On the other hand however, quantum effects in 4D dilatonic
black hole are -- to the best of my knowledge -- practically unexplored.
This does not mean that this group of problems is uninteresting: belonging
to the realm of the low-energy approximation to string theory ($a=1)$ or the
Kaluza-Klein theory ($a=\sqrt{3}$), the dilatonic black holes would interact
with various quantized fields. The main obstacle preventing construction of
the renormalized stress-energy tensor is the computational complexity of the
problem.

The evaporation process of the massless scalar field noninteracting with a
dilaton field has been analysed in \cite{koga1,cqg} whereas the field
fluctuation, $\langle \phi ^{2}\rangle ,$ of the minimally coupled massless
scalar field in the vicinity of the event horizon of the dilatonic black
hole has been studied by Shiraishi \cite{kyioshi}. Specifically, it was
shown that the emission rate of the Hawking radiation blows up near the
extremality limit for $a>1.$ On the other hand it is finite for $a=1$ and
zero for $a<1.$ The field fluctuation diverges for $a>0$ for the extremal
configuration.

\section{The geometry}

Functionally differentiating $S$ with respect to the metric tensor, dilaton
field, and Maxwell field one obtains the system of equations of motion that
could be solved exactly. Static and spherically-symmetric solution has been
found by Gibbons and Maeda~\cite{gibbons2}, and by Garfinkle, Horowitz and
Strominger~\cite{horowitz}: 
\begin{equation}
ds^{2}=A\left( r\right) dt^{2}+%
%TCIMACRO{\dfrac{dr^{2}}{A\left( r\right) }}%
%BeginExpansion
{\displaystyle{dr^{2} \over A\left( r\right) }}%
%EndExpansion
+B^{2}\left( r\right) d\Omega ^{2},  \label{el1}
\end{equation}
where 
\begin{equation}
A\left( r\right) =\left( 1-%
%TCIMACRO{\dfrac{r_{+}}{r}}%
%BeginExpansion
{\displaystyle{r_{+} \over r}}%
%EndExpansion
\right) \left( 1-%
%TCIMACRO{\dfrac{r_{-}}{r}}%
%BeginExpansion
{\displaystyle{r_{-} \over r}}%
%EndExpansion
\right) ^{%
%TCIMACRO{\tfrac{1-a^{2}}{1+a^{2}}}%
%BeginExpansion
{\textstyle{1-a^{2} \over 1+a^{2}}}%
%EndExpansion
}  \label{el2}
\end{equation}
and 
\begin{equation}
B^{2}=r^{2}\left( 1-%
%TCIMACRO{\dfrac{r_{-}}{r}}%
%BeginExpansion
{\displaystyle{r_{-} \over r}}%
%EndExpansion
\right) ^{%
%TCIMACRO{\tfrac{2a^{2}}{1+a^{2}}}%
%BeginExpansion
{\textstyle{2a^{2} \over 1+a^{2}}}%
%EndExpansion
}.  \label{el3}
\end{equation}
The integration constants $r_{+}$ and $r_{-}$ are related to the mass and
charge of the dilatonic black hole according to 
\begin{equation}
2M=r_{+}+\left( 
%TCIMACRO{\dfrac{1-a^{2}}{1+a^{2}}}%
%BeginExpansion
{\displaystyle{1-a^{2} \over 1+a^{2}}}%
%EndExpansion
\right) r_{-}  \label{rr1}
\end{equation}
and 
\begin{equation}
Q^{2}=%
%TCIMACRO{\dfrac{r_{+}r_{-}}{1+a^{2}} }%
%BeginExpansion
{\displaystyle{r_{+}r_{-} \over 1+a^{2}}}.%
%EndExpansion
\label{rr2}
\end{equation}
The dilaton field is given by 
\begin{equation}
e^{2a\phi }=\left( 1-%
%TCIMACRO{\dfrac{r_{-}}{r}}%
%BeginExpansion
{\displaystyle{r_{-} \over r}}%
%EndExpansion
\right) ^{%
%TCIMACRO{\tfrac{2a^{2}}{1+a^{2}}}%
%BeginExpansion
{\textstyle{2a^{2} \over 1+a^{2}}}%
%EndExpansion
},
\end{equation}
whereas the electric field is simply $F=%
%TCIMACRO{\dfrac{Q}{r^{2}}}%
%BeginExpansion
{\displaystyle{Q \over r^{2}}}%
%EndExpansion
dt\wedge dr.$ Inspection of the line element shows that the event horizon is
located at $r_{+};$ at $r=$ $r_{-}$ one has a coordinate singularity that
could be ignored so long one considers region $r\geq r_{+}>r_{-}.$ The
choice $a=0$ leads to the Reissner-Nordstr\"{o}m solution. At $\left|
Q\right| /M=\left( 1+a^{2}\right) ^{1/2}$, a case usually addressed to as an
extremal black hole, the event horizon and $r_{-}$ coincides and in this
limit the surface $r=r_{+}$ $=r_{-}$ is zero except $a=0.$ Although more
realistic models require massive $\phi $ field, the dilatonic solutions (\ref
{el1}-\ref{el3}) are of principal interest as they provide useful models for
studies of the consequences of modifications of the geometries of the
classical black holes. Finally, observe that the Kretschmann scalar $K$
computed at the event horizon near the extremality limit behaves as 
\begin{equation}
K\sim \left( r_{+}-r_{-}\right) ^{-%
%TCIMACRO{\tfrac{4a^{2}}{1+a^{2}}}%
%BeginExpansion
{\textstyle{4a^{2} \over 1+a^{2}}}%
%EndExpansion
}.
\end{equation}

\section{The renormalized stress-energy tensor}

\subsection{Approximate effective action}

In the framework of the Schwinger-DeWitt approximation the first order
effective action of the massive scalar field is constructed
from the coincidence limit of the coefficient $a_{3}\left( x,x^{\prime
}\right) .$ Inserting $\left[ a_{3}\right] $ as given in Appendix into (\ref
{weff}), integrating by parts and finally making use of the elementary
properties of the Riemann tensor, after necessary simplifications one has 
\cite{AvraPhD,Avraros,Avranucl}: 
\begin{eqnarray}
W_{ren}^{(1)}\, &=&\frac{1}{192\pi ^{2}m^{2}}\int d^{4}x\sqrt{g}\left[ {{%
\frac{1}{2}}}\left( \eta ^{2}-{{\frac{\eta }{15}}}-{{\frac{1}{315}}}\right)
R\Box R\,+\,{\frac{1}{140}}R_{pq}\Box R^{pq}-{\eta }^{3}R^{3}\,+\,{\frac{1}{%
30}\eta }RR_{pq}R^{pq}\,-\,{\frac{1}{30}\eta }RR_{pqab}R^{pqab}\right. 
\nonumber \\
&&\left. -{\frac{8}{945}}R_{q}^{p}R_{a}^{q}R_{p}^{a}\,+\,{\frac{2}{315}}%
R^{pq}R_{ab}R_{~p~q}^{a~b}\,+\,\,{\frac{1}{1260}}R_{pq}R_{~cab}^{p}R^{qcab}+{%
\frac{17}{7560}}{R_{ab}}^{pq}{R_{pq}}^{cd}{R_{cd}}^{ab}-\,{\frac{1}{270}}%
R_{~p~q}^{a~b}R_{~c~d}^{p~q}R\right]  \nonumber \\
&=&{{\frac{1}{192\pi ^{2}m^{2}}}}\sum_{i=1}^{10}\alpha _{i}W_{\left(
i\right) },  \label{ww}
\end{eqnarray}
were $\eta =\xi -1\slash6$ and $\alpha _{i\text{ }}$ are equal to the
numerical coefficients that stand in front of the geometrical terms in the
right hand side of the equation (\ref{ww}).

Differentiating functionally $W_{ren}^{\left( 1\right) }$ with respect to a
metric tensor one obtains rather complicated expression which schematically
may be written as 
\begin{equation}
\,\langle T^{ab}\rangle =\sum_{i=1}^{10}\alpha _{i}\tilde{T}^{\left(
i\right) ab}={{\frac{1}{96\pi ^{2}m^{2}\sqrt{g}}}}\sum_{i=1}^{10}\alpha _{i}{%
{\frac{\delta W_{\left( i\right) }}{\delta g_{ab}}}}=T^{\left( 0\right)
ab}+\eta T^{\left( 1\right) ab}+\eta ^{2}T^{\left( 2\right) ab}+\eta
^{3}T^{\left( 3\right) ab},  \label{ten}
\end{equation}
where each $\tilde{T}^{\left( i\right) ab}$ is constructed solely from the
curvature tensor, its covariant derivatives and contractions. Because of the
complexity of the resulting stress-energy tensor it will be not presented
here and for its full form as well as the technical details the reader is
referred to \cite{kocio1,kocio2}. The result may be easily extended to
fields of other spins as the appropriate tensors differ by numerical
coefficients $\alpha _{i}$ only.

The coincidence limit of $a_{4}\left( x,x^{\prime }\right) $ is also known:
it has been calculated by Avramidi~\cite{AvraPhD,Avraros,Avranucl} and by
Amsterdamski, Berkin and O'Connor~\cite{Amsterdamski}. In principle, the
above procedure could be extended to include $m^{-4}$ terms and the general
structure of $\left[ a_{4}\right] $ indicates that the second-order
stress-energy tensor divides naturally into five terms $\sum_{i=0}^{4}\eta
^{i}T^{\left( i\right) ab}.$ Unfortunately, since the effective action
constructed from $\left[ a_{4}\right] $ is extremely complicated, so is its
functional derivative and the practical use of the thus obtained result may be
a real challenge. However, $\left[ a_{4}\right] $ still could be employed in
the analyses of the field fluctuation.

In order to simplify our discussion let us define $q=\left| Q\right| /M$, $%
x_{\pm }=r_{\pm }/M$ and $x=r/M$. The Schwinger-DeWitt technique may be used
when the characteristic radius of curvature in much greater than the Compton
length. Simple considerations indicate that for $r\gg r_{+}$ $\ $it could be
used for arbitrary value of $a.$ Assuming that ${\Bbb L}$ is related to the
Kretschmann scalar as 
\begin{equation}
R_{a b c d} R^{a b c d} \sim {\Bbb L}^{-4}
\end{equation}
the condition of applicability of the approximation near the event horizon
could be written as 
\begin{equation}
%TCIMACRO{\dfrac{2c}{M^{2}x_{+}^{3}}}%
%BeginExpansion
{\displaystyle{2c \over M^{2}x_{+}^{3}}}%
%EndExpansion
\left( x_{+}-x_{-}\right) ^{-%
%TCIMACRO{\tfrac{2a^{2}}{1+a^{2}}}%
%BeginExpansion
{\textstyle{2a^{2} \over 1+a^{2}}}%
%EndExpansion
}\ll m^{2},  \label{cond1}
\end{equation}
where $c^{2}=2x_{-}^{2}+\left[ x_{-}-\left( 1+a^{2}\right) x_{+}\right]
^{2}. $ It is evident that for $a>0$ the Schwinger-DeWitt approximation is
inapplicable for $r_{+}$ close to $r_{-}.$ For the extremal
Reissner-Nordstr\"{o}m black hole this condition becomes $M^{2}m^{2}\gg 2%
\sqrt{2}.$

The temperature of the dilatonic black hole obtained by means of standard
methods is given by 
\begin{mathletters}
\begin{equation}
T_{H}=%
%TCIMACRO{\dfrac{1}{4\pi Mx_{+}}}%
%BeginExpansion
{\displaystyle{1 \over 4\pi Mx_{+}}}%
%EndExpansion
\left( 1-%
%TCIMACRO{\dfrac{x_{-}}{x_{+}}}%
%BeginExpansion
{\displaystyle{x_{-} \over x_{+}}}%
%EndExpansion
\right) ^{%
%TCIMACRO{\tfrac{1-a^{2}}{1+a^{2}}}%
%BeginExpansion
{\textstyle{1-a^{2} \over 1+a^{2}}}%
%EndExpansion
}  \label{temp}
\end{equation}
and for given $q$ it depends on the dilatonic coupling. Inspection of (\ref
{temp}) shows that 
\end{mathletters}
\begin{eqnarray*}
T_{H} &<&(8\pi M)^{-1}\qquad \left( a<1\right) , \\
T_{H} &=&\left( 8\pi M\right) ^{-1}\qquad \left( a=1\right) , \\
T_{H} &>&\left( 8\pi M\right) ^{-1}\qquad \left( a>1\right) .
\end{eqnarray*}
The temperature of the extremal configuration is zero for $a<1,$ takes the
same value as for a Schwarzschild black hole for $a=1,$ and diverges for $%
a>0.$ Moreover, it is easily seen that the condition $T_{H}\ll m$ is
violated for $a>1$ near the extremality limit.

\subsection{General case}

Solving the system (\ref{rr1}) and (\ref{rr2}) one easily obtains 
\begin{equation}
x_{+}=1+\sqrt{1-\left( 1-a^{2}\right) q^{2}}
\end{equation}
and 
\begin{equation}
x_{-}=%
%TCIMACRO{\dfrac{1+a^{2}}{1-a^{2}}}%
%BeginExpansion
{\displaystyle{1+a^{2} \over 1-a^{2}}}%
%EndExpansion
\left( 1-\sqrt{1-\left( 1-a^{2}\right) q^{2}}\right) .
\end{equation}
Before proceeding further let us observe that $R=0$ for $a=0,$ and,
consequently, $\delta W_{\left( 1\right) }/\delta g_{ab}$ and $\delta
W_{\left( 3\right) }/\delta g_{ab}$ is zero. The stress-energy tensor has
therefore a simple form 
\begin{equation}
\langle T_{a}^{b}\rangle ^{a=0}=T_{a}^{\left( 0\right) b}+\eta T_{a}^{\left(
1\right) b}.  \label{setrn}
\end{equation}
On the other hand, the curvature scalar vanishes at the event horizon for
any $a$ and is ${\cal O}\left( q^{4}\right) $ for small $q$ elsewhere.
Moreover, since $\partial _{r}R$ is the only nonzero component of $\nabla
_{a}R$ one concludes that $T_{a}^{\left( 3\right) b}\left( r_{+}\right) =0$
and is negligible in the closest vicinity of $r_{+}.$ It is because the only
nonvanishing in this limit term is proportional to 
\begin{equation}
\nabla _{a}R\nabla ^{b}R-\left( \nabla R\right) ^{2}\delta _{a}^{b}.
\end{equation}
A closer examination indicates that $T_{a}^{\left( 3\right) b}$ is $O\left(
q^{8}\right) .$ Similarly, one expects that for small $q$ the term $%
T_{a}^{\left( 2\right) b}$ is of order $O\left( q^{4}\right) .$ On the other
hand, the contribution of the last two terms in the right hand side of
equation (\ref{ten}) could be made arbitrarily large by a suitable choice of
the curvature coupling. It should be noted however that such values of $\eta 
$ are clearly unphysical and should be rejected.

Restricting to the exterior region and calculating components of the Riemann
tensor, its contractions and covariant derivatives to the required order,
after some algebra one arrives at the rather complicated result, that for
obvious reasons will not be presented here. However it could be
schematically written in surprisingly simple form 
\begin{equation}
\langle T_{a}^{b}\rangle =%
%TCIMACRO{\dfrac{p}{\left( 1+a^{2}\right) \,x^{15}}}%
%BeginExpansion
{\displaystyle{p \over \left( 1+a^{2}\right) \,x^{15}}}%
%EndExpansion
\left( 1-%
%TCIMACRO{\dfrac{x_{-}}{x}}%
%BeginExpansion
{\displaystyle{x_{-} \over x}}%
%EndExpansion
\right) ^{-%
%TCIMACRO{\tfrac{3\left( 3a^{2}+1\right) }{1+a^{2}}}%
%BeginExpansion
{\textstyle{3\left( 3a^{2}+1\right)  \over 1+a^{2}}}%
%EndExpansion
}\sum_{ijk}d_{ijkb}^{a}\left[ \eta ,\,a^{2}\right] \,x^{i}x_{+}^{j}x_{-}^{k}
\label{schema}
\end{equation}
with $0\leq i\leq 7$, $0\leq j\leq 3$ and $0\leq k\leq 6$ subjected to the
condition $i+j+k=9.$ Here 
\begin{equation}
p=%
%TCIMACRO{\dfrac{1}{192\pi ^{2}m^{2}M^{6}}}%
%BeginExpansion
{\displaystyle{1 \over 192\pi ^{2}m^{2}M^{6}}}%
%EndExpansion
.
\end{equation}
and $d_{ijkb}^{a}$ for given $a$ and $\eta $ are numerical coefficients.
Some extra work shows that the tensor (\ref{schema}) is covariantly
conserved and is regular for regular geometries. Moreover, the difference $%
\langle T_{t}^{t}\rangle -\langle T_{r}^{r}\rangle $ factors 
\begin{equation}
\langle T_{t}^{t}\rangle -\langle T_{r}^{r}\rangle =%
%TCIMACRO{\dfrac{p}{\left( 1+a^{2}\right) \,x^{14}}}%
%BeginExpansion
{\displaystyle{p \over \left( 1+a^{2}\right) \,x^{14}}}%
%EndExpansion
\left( 1-%
%TCIMACRO{\dfrac{x_{+}}{x}}%
%BeginExpansion
{\displaystyle{x_{+} \over x}}%
%EndExpansion
\right) \left( 1-%
%TCIMACRO{\dfrac{x_{-}}{x}}%
%BeginExpansion
{\displaystyle{x_{-} \over x}}%
%EndExpansion
\right) ^{-%
%TCIMACRO{\tfrac{3\left( 3a^{2}+1\right) }{1+a^{2}}}%
%BeginExpansion
{\textstyle{3\left( 3a^{2}+1\right)  \over 1+a^{2}}}%
%EndExpansion
}f\left( x\right) ,  \label{fact}
\end{equation}
where the regular function $f\sim (x_{+}-x_{-})^{2}$ as $x_{-}\rightarrow
x_{+}$ and consequently within the domain of applicability of the
Schwinger-DeWitt approximation the stress-energy tensor is regular in a
freely falling frame.

It could be demonstrated by that the trace of the stress-energy tensor of
the massive scalar field has a simple form 
\begin{equation}
\langle T_{a}^{a}\rangle \,=\,{\frac{1}{16\pi ^{2}m^{2}}}\left\{ 3\left( \xi
-{\frac{1}{6}}\right) \Box \lbrack a_{2}]\,-\,[a_{3}]\right\} .  \label{slad}
\end{equation}
This equation together with 
\begin{equation}
\nabla _{b}T_{a}^{b}=0  \label{covcons}
\end{equation}
may serve as an independent check of the calculations. For conformally
coupled fields the trace is proportional to the coincidence limit of $%
[a_{3}]. $ We remark here that for conformally invariant massless scalar
field the anomalous trace is proportional to $[a_{2}];$ it should be noted
however, that (\ref{slad}) has been calculated for $\langle T_{a}^{b}\rangle 
$ given by (\ref{ten}) whereas the trace of the conformally invariant
massless fields is a general property of the regularized stress-energy
tensor.

Since the practical use of the general result is severely limited, it is
instructive to analyse the stress-energy tensor in some specific cases. In
the latter we shall confine our analysis to $0\,\leq a\,\leq \sqrt{3}$ with
the special emphasis put on the case $a=1.$ However, before proceeding to
examination of some concrete choices of $a$ let us analyse $\langle
T_{a}^{b}\rangle $ for small $q.$

\subsection{Arbitrary $a,$ $q\ll 1$}

Assuming $q\ll 1$, expanding $\langle T_{a}^{b}\rangle $ into a power series,
and finally collecting the terms with the like powers of $q$ one has

\begin{equation}
\,\langle T_{a}^{b}\rangle =\,\langle T_{a}^{b}\rangle ^{a=0}+%
%TCIMACRO{\dfrac{a^{2}}{96\pi ^{2}m^{2}x^{10}M^{6}}}%
%BeginExpansion
{\displaystyle{a^{2} \over 96\pi ^{2}m^{2}x^{10}M^{6}}}%
%EndExpansion
(q^{2}t_{a}^{\left( 1\right) b}+q^{4}t_{a}^{\left( 2\right)
b}+q^{6}t_{a}^{\left( 3\right) b}+...),  \label{ser1}
\end{equation}
where $\langle T_{b}^{b}\rangle ^{a=0}$ is evaluated for $a=0$ and coincides
with the expression describing the stress-energy tensor in the geometry of
the Reissner-Nordstr\"{o}m black hole~\cite{AHS,kocio1}. The explicit
expressions for the coefficients $t_{a}^{\left( i\right) b}$ as well as the
components of $\langle T_{b}^{b}\rangle ^{a=0}$ are listed in the appendix.
A closer examination shows that for $q\lesssim 0.7$ the expansion (\ref{ser1}%
) reproduces the general result satisfactorily, and, moreover, for $%
q\lesssim 1/3$ the results weakly depend on the coupling $a.$ From (\ref
{ser1}) it is evident that for $a=0$ and $q=0$ the stress-energy tensor
reduces to the expression derived by Frolov and Zel'nikov in the geometry of
the Schwarzschild black hole~\cite{fz1,FrolovNovikov}.

\subsection{Dilatonic black hole $a=1$}

In this subsection we shall construct and investigate the stress-energy
tensor of the massive scalar field resulting from (\ref{schema}) for the
particular combinations of couplings. Consider $a=1.$ Since the second
factor in $A\left( r\right) $ vanishes, we expect considerable
simplifications as the event horizon is now located at $2M$ whereas the
`inner' one at $q^{2}M$. Indeed, defining $y=r/r_{+},$ equation (\ref{schema}%
) could be written in a simple form: 
\begin{equation}
\langle T_{a}^{b}\rangle =%
%TCIMACRO{\dfrac{p}{\left( 2y-q^{2}\right) ^{6}}}%
%BeginExpansion
{\displaystyle{p \over \left( 2y-q^{2}\right) ^{6}}}%
%EndExpansion
\sum_{ij}b_{ija}^{b}\left[ \eta \right] \,q^{2i}y^{-j-2}  \label{w1}
\end{equation}
with $0\leq i\leq 6$ and $0\leq j\leq 7,$ where $b_{ija}^{b}$ are numerical
coefficients. From (\ref{fact}) it could be shown that for any $\eta $ the
difference $\langle T_{t}^{t}\rangle -\langle T_{r}^{r}\rangle $ factorizes
as 
\begin{equation}
\langle T_{t}^{t}\rangle -\langle T_{r}^{r}\rangle =%
%TCIMACRO{\dfrac{1-y}{y^{9}\left( q^{2}-2y\right) ^{6}}}%
%BeginExpansion
{\displaystyle{1-y \over y^{9}\left( q^{2}-2y\right) ^{6}}}%
%EndExpansion
f\left( y\right) ,
\end{equation}
where, for $0\leq q<\sqrt{2}$ the function $f\left( y\right) $ is regular at
the event horizon. Equation (\ref{w1}) could be contrasted to the analogous
expression evaluated in the Reissner-Nordstr\"{o}m geometry $\left(
a=0\right) :$ 
\begin{equation}
\langle T_{a}^{b}\rangle =\frac{p}{y^{6}}\sum_{ij}c_{ija}^{b}\left[ \eta %
\right] q^{2i}y^{-j-2},  \label{w2}
\end{equation}
where $0\leq i\leq 3$, $0\leq j\leq 4$, and $c_{ija}^{b}$ are another set of
numerical coefficients.

To perform quantitative analysis however, we have to refer to exact
formulas. For $\eta =0$ it suffices to compute only $T_{a}^{\left( 0\right)
b}$ as the others terms do not contribute to the final result. After some
algebra one has

%%% konforemne sprzezenie
%% Ttt
\begin{eqnarray}
\langle T_{t}^{t}\rangle &=&%
%TCIMACRO{\dfrac{p}{\left( 2y-q^{2}\right) ^{6}}}%
%BeginExpansion
{\displaystyle{p \over \left( 2y-q^{2}\right) ^{6}}}%
%EndExpansion
\left[ \frac{313}{210\,y^{3}}-\frac{19}{14\,y^{2}}-q^{2}\,\left( \frac{61}{%
30\,y^{4}}-\frac{31}{70\,y^{3}}-\frac{9}{7\,y^{2}}\right) +q^{4}\,\left( 
\frac{143}{840\,y^{5}}+\frac{7313}{2520\,y^{4}}-\frac{577}{210\,y^{3}}-\frac{%
1}{28\,y^{2}}\right) \right.  \nonumber \\
&&+q^{6}\,\left( \frac{1381}{1120\,y^{6}}-\frac{6607}{1680\,y^{5}}+\frac{1813%
}{720\,y^{4}}+\frac{1}{28\,y^{3}}\right) -q^{8}\,\left( \frac{9277}{%
10080\,y^{7}}-\frac{43837}{20160\,y^{6}}+\frac{1007}{840\,y^{5}}+\frac{139}{%
10080\,y^{4}}\right)  \nonumber \\
&&\left. +q^{10}\,\left( \frac{1817}{6720\,y^{8}}-\frac{479}{840\,y^{7}}+%
\frac{559}{1920\,y^{6}}+\frac{7}{2880\,y^{5}}\right) -q^{12}\,\left( \frac{%
1783}{60480\,y^{9}}-\frac{473}{8064\,y^{8}}+\frac{11}{384\,y^{7}}+\frac{1}{%
6912\,y^{6}}\right) \right]  \nonumber \\
&&  \label{dilconf1}
\end{eqnarray}

%%Trr
\begin{eqnarray}
\langle T_{r}^{r}\rangle &=&%
%TCIMACRO{\dfrac{p}{\left( 2y-q^{2}\right) ^{6}}}%
%BeginExpansion
{\displaystyle{p \over \left( 2y-q^{2}\right) ^{6}}}%
%EndExpansion
\left[ \frac{1}{2\,y^{2}}-\frac{11}{30\,y^{3}}+q^{2}\,\left( \frac{47}{%
42\,y^{4}}-\frac{353}{210\,y^{3}}+\frac{9}{35\,y^{2}}\right) -q^{4}\,\left( 
\frac{481}{280\,y^{5}}-\frac{7267}{2520\,y^{4}}+\frac{43}{45\,y^{3}}-\frac{11%
}{140\,y^{2}}\right) \right.  \nonumber \\
&&+q^{6}\,\left( \frac{895}{672\,y^{6}}-\frac{11687}{5040\,y^{5}}+\frac{103}{%
112\,y^{4}}-\frac{11}{140\,y^{3}}\right) -q^{8}\,\left( \frac{61}{112\,y^{7}}%
-\frac{2143}{2240\,y^{6}}+\frac{2053}{5040\,y^{5}}-\frac{53}{1440\,y^{4}}%
\right)  \nonumber \\
&&\left. +q^{10}\,\left( \frac{2269}{20160\,y^{8}}-\frac{397}{2016\,y^{7}}+%
\frac{233}{2688\,y^{6}}-\frac{173}{20160\,y^{5}}\right) -q^{12}\,\left( 
\frac{139}{15120\,y^{9}}-\frac{91}{5760\,y^{8}}+\frac{1}{144\,y^{7}}-\frac{5%
}{6912\,y^{6}}\right) \right]  \nonumber \\
&&
\end{eqnarray}

%%Tang
and 
\begin{eqnarray}
\langle T_{\theta }^{\theta }\rangle &=&%
%TCIMACRO{\dfrac{p}{\left( 2y-q^{2}\right) ^{6}}}%
%BeginExpansion
{\displaystyle{p \over \left( 2y-q^{2}\right) ^{6}}}%
%EndExpansion
\left[ \frac{367}{210\,y^{3}}-\frac{3}{2\,y^{2}}-q^{2}\,\left( \frac{367}{%
70\,y^{4}}-\frac{1129}{210\,y^{3}}+\frac{27}{35\,y^{2}}\right)
+q^{4}\,\left( \frac{65}{8\,y^{5}}-\frac{4379}{420\,y^{4}}+\frac{146}{%
45\,y^{3}}-\frac{33}{140\,y^{2}}\right) \right.  \nonumber \\
&&-q^{6}\,\left( \frac{11099}{1680\,y^{6}}-\frac{11749}{1260\,y^{5}}+\frac{%
1423}{420\,y^{4}}-\frac{33}{140\,y^{3}}\right) +q^{8}\,\left( \frac{59011}{%
20160\,y^{7}}-\frac{29209}{6720\,y^{6}}+\frac{1907}{1120\,y^{5}}-\frac{643}{%
5040\,y^{4}}\right)  \nonumber \\
&&\left. -q^{10}\,\left( \frac{13589}{20160\,y^{8}}-\frac{6943}{6720\,y^{7}}+%
\frac{8551}{20160\,y^{6}}-\frac{173}{5040\,y^{5}}\right) +q^{12}\,\left( 
\frac{7669}{120960\,y^{9}}-\frac{143}{1440\,y^{8}}+\frac{97}{2304\,y^{7}}-%
\frac{25}{6912\,y^{6}}\right) \right]  \nonumber \\
&&  \label{dilconf3}
\end{eqnarray}
In fact it suffices to know only one component of the stress-energy tensor,
say $\langle T_{\theta }^{\theta }\rangle ,$ as the remaining ones could be
easily obtained solving equations (\ref{slad}) and (\ref{covcons}) and
putting the integration constant to zero.

Despite its similarity with the Schwarzschild line element, the nonextremal $%
a=1$ dilatonic \ black holes have much in common with the
Reissner-Nordstr\"{o}m solution. We shall, therefore, address the question
of how the differences between the geometry of the Reissner-Nordstr\"{o}m
black hole on the one hand and the dilatonic black hole on the other are
reflected in the overall behaviour of our approximate stress-energy tensors.
First, from (\ref{dilconf1}-\ref{dilconf3}) it could be easily inferred that 
$\langle T_{a}^{b}\rangle $ evaluated for the extremal configuration is
divergent as $y\rightarrow 1$. Indeed, for $q=\sqrt{2}$ the components of
the stress-energy tensor behave as $\left( y-1\right) ^{-3}.$ This is in a
sharp contrast with the Reissner-Nordstr\"{o}m case, in which the
stress-energy tensor approaches 
\begin{equation}
\langle T_{a}^{b}\rangle =%
%TCIMACRO{\dfrac{1}{2880\pi ^{2}m^{2}M^{6}}}%
%BeginExpansion
{\displaystyle{1 \over 2880\pi ^{2}m^{2}M^{6}}}%
%EndExpansion
\left[ 
%TCIMACRO{\dfrac{16}{21}}%
%BeginExpansion
{\displaystyle{16 \over 21}}%
%EndExpansion
-\left( \xi -%
%TCIMACRO{\dfrac{1}{6}}%
%BeginExpansion
{\displaystyle{1 \over 6}}%
%EndExpansion
\right) \right] {\rm diag}\left[ 1,1,-1,-1\right]
\end{equation}
as $y\rightarrow 1$. It should be noted however, that, except $a=0,$ the
region in the vicinity of the degenerate horizon of the extremal geometry is
beyond the applicability of the Schwinger-DeWitt approximation. On the other
hand, however, one expects that in the opposite limit, i.e. for $q\ll 1$,
the appropriate components of the stress-energy tensor are almost
indistinguishable.

To analyse $\langle T_{a}^{b}\rangle $ for intermediate values of $q$ let us
refer to the numerical calculations. The plots of the time radial and
angular components of the stress-energy tensor of the quantized massive
scalar field as a function of the rescaled radial coordinate for five
exemplar values of $q=\,$0.1, 0.2, 0.5, 0.65 and 0.85 are displayed in
figures 1--3. Inspection of the figures and comparison with the analogous
results obtained for the Reissner-Nordstr\"{o}m geometry indicates that even
for the intermediate values of $q$ there are still qualitative similarities.
Indeed, the time and angular components attain (positive) maximum at the
event horizon, decrease with r and approach (negative) minimum. The
magnitude of the maximum and the modulus of the minimum increase with
increasing $q,$ and, consequently, so does the slope of the curves. 

Before proceeding to physically interesting and important case $\eta =-1/6$,
it is useful to study a role played by each $T_{a}^{\left( i\right) b}$
separately. First, $\,$it could be easily shown that $T_{a}^{\left( 3\right)
b}$ is negligible with respect to other terms, and, therefore, it does not
contribute to the final result for reasonable values of the curvature
coupling. The run of the resulting stress-energy tensor depends on a
competition between remaining components. Indeed, inspection of figure 4 
in which we exhibited $T_{t}^{\left( i\right) t}$ as a function of the
rescaled radial coordinate for four exemplar values of $q$ indicates that
the term $-T_{a}^{\left( 1\right) b}$ produces the most prominent maximum at
the event horizon for $q\lesssim 0.9$ whereas for greater values of $q$ this
role is played by $T_{a}^{\left( 2\right) b}.$ General features of $%
T_{r}^{\left( i\right) r}$ and $T_{\theta }^{\left( i\right) \theta }$ are
essentially the same.

%%%%%%%%%%%%%%%%%%%%%%%%%%%%%%%%%%%%%%%%%%%%%%%%%%%%%%%%%%%%%%%%%%%%%%%%%%%
%%%%%%%%%%%% ********** rysunek1  **********************
%%%% ******************** figure 1 ****************************************
\begin{figure}[tbh]
\centering
\includegraphics[width=12.5cm]{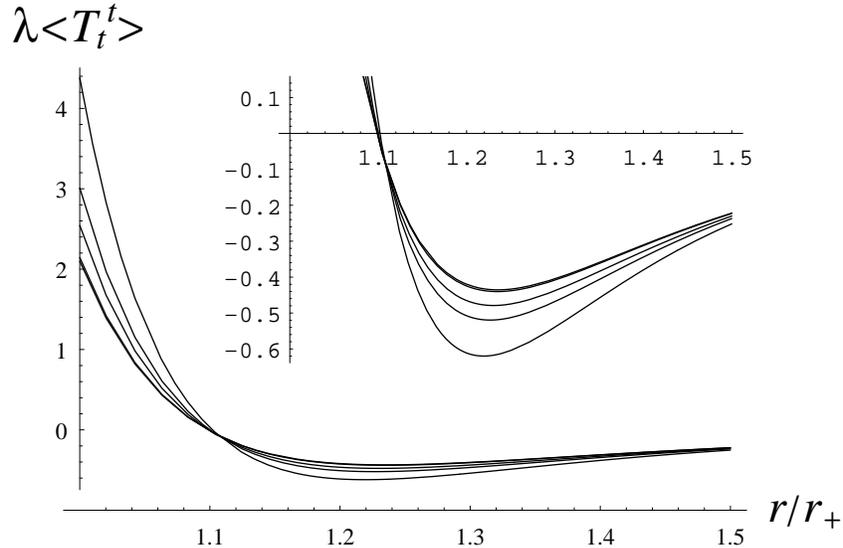}
\caption[ttt]{This graph shows the radial dependence of the rescaled
component $\protect\lambda \langle T_{t}^{t}\rangle ,$ $(\protect\lambda
=10^{3}/p)$ of the stress-energy tensor of the massive conformally coupled
scalar field in the geometry of the dilatonic black hole with $a=1.$ From
top to bottom at the event horizon the curves are for $|Q|/M$ = 0.1, 0.2,
0.5, 0.65 and 0.85. In each case $\langle T_{t}^{t}\rangle $ has its
positive maximum at $r_{+}$ and attains negative minimum (right panel) away
from the event horizon. }
\label{fig1}
\end{figure}
%%%%%%%%%%%%%%%%%%%%%%%%%%%%%%%%%%%%%%%%%%%%%%%%%%%%%%%%%%%%%%%%%%%%%%%%%%%
%%%%%%%%%%%%% ********* rysunek2 ***************
%%% ********************* figure 2 *****************************************
\begin{figure}[tbh]
\centering
\includegraphics[width=12.5cm]{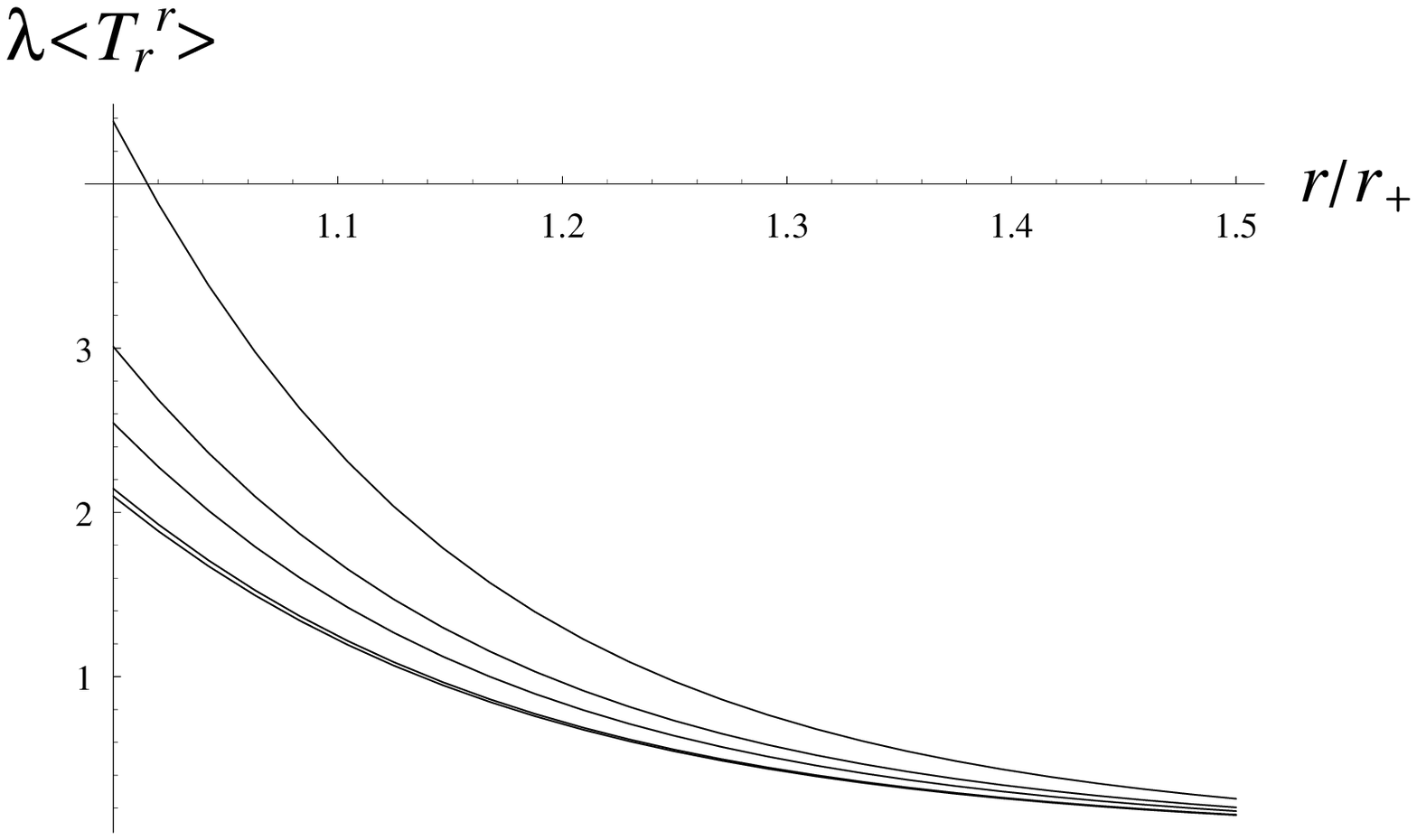}
\caption[rrrr]{This graph shows the radial dependence of the rescaled
component $\protect\lambda \langle T_{r}^{r}\rangle ,$ $(\protect\lambda
=10^{3}/p)$of the stress-energy tensor of the massive conformally coupled
scalar field in the geometry of the dilatonic black hole with $a=1.$ From
top to bottom at the event horizon the curves are for $|Q|/M$ = 0.1, 0.2,
0.5, 0.65 and 0.85. In each case $\langle T_{tr}^{r}\rangle $ has its
positive maximum at $r_{+}$ and monotonically decreases with r.}
\label{fig2}
\end{figure}

Now the run of the stress-energy tensor as a function of $q$ could be easily
anticipated. The general structure remains, of course, of the form (\ref{w1}%
), but now the dominant contribution to the result is provided initially by
the term $\eta T_{a}^{\left( 1\right) b}$ and subsequently with increasing $%
q $ by the sum $-1/6\,T_{a}^{\left( 1\right) b}+1/36\,T_{a}^{\left( 2\right)
b}.$ Moreover, since oscillatory-like behavior of $T_{a}^{\left( 3\right) b}$
does not play a significant role we have also qualitative similarities with
the tensor evaluated for the conformal coupling.

%%%%%%%%%%%%%%%%%%%%%%%%%%%%%%%%%%%%%%%%%%%%%%%%%%%%%%%%%%%%%%%%%%%%%%%%%%%
%%%%%%%%%%%%% ********* rysunek3 ***************
%%% ********************* figure 3 *****************************************
\begin{figure}[tbh]
\centering
\includegraphics[width=12.5cm]{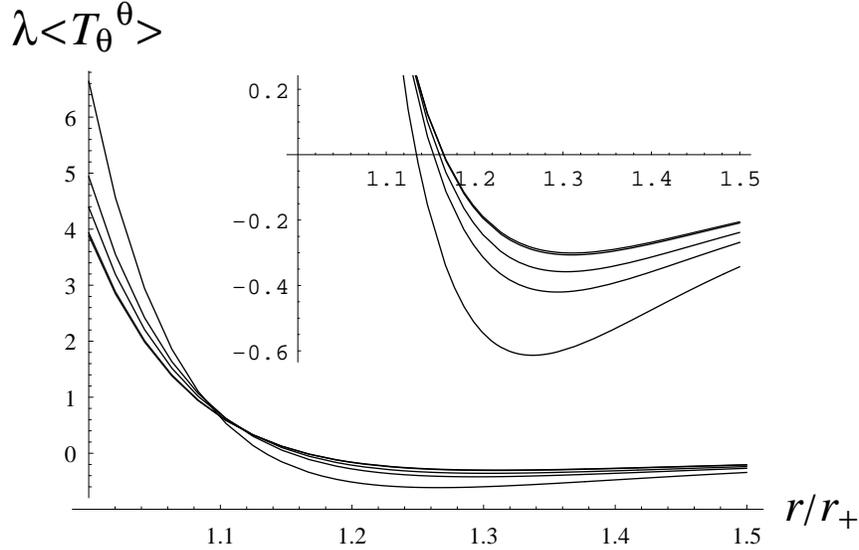}
\caption[aaa]{This graph shows the radial dependence of the rescaled
component $\protect\lambda \langle T_{\protect\theta }^{\protect\theta
}\rangle ,$ $(\protect\lambda =10^{3}/p)$ of the stress-energy tensor of the
massive conformally coupled scalar field in the geometry of the dilatonic
black hole with $a=1.$ From top to bottom at the event horizon the curves
are for $|Q|/M$ = 0.1, 0.2, 0.5, 0.65 and 0.85. In each case $\langle
T_{t}^{t}\rangle $ has its positive maximum at ${+}$ and and attains
negative minimum (right panel) away from the event horizon.}
\label{fig3}
\end{figure}
%%%%%%%%%%%%%%%%%%%%%%%%%%%%%%%%%%%%%%%%%%%%%%%%%%%%%%%%%%%%%%%%%%%%%%%%%%%

Having computed $\tilde{T}_{a}^{\left( i\right) b}$ and combining them with
appropriate values of the coefficients $\alpha _{i}$ for $i=1..10,$ after
simplifications and rearrangement one has 
\begin{eqnarray}
\langle T_{t}^{t}\rangle &=&%
%TCIMACRO{\dfrac{p}{\left( 2y-q^{2}\right) ^{6}}}%
%BeginExpansion
{\displaystyle{p \over \left( 2y-q^{2}\right) ^{6}}}%
%EndExpansion
\left[ \frac{1237}{210\,y^{3}}-\frac{75}{14\,y^{2}}-q^{2}\,\left( \frac{377}{%
30\,y^{4}}-\frac{703}{70\,y^{3}}-\frac{9}{7\,y^{2}}\right) +q^{4}\,\left( 
\frac{3637}{140\,y^{5}}-\frac{5219}{140\,y^{4}}+\frac{2259}{140\,y^{3}}-%
\frac{99}{28\,y^{2}}\right) \right.  \nonumber \\
&&-q^{6}\,\left( \frac{84407}{3360\,y^{6}}-\frac{11311}{280\,y^{5}}+\frac{%
1177}{60\,y^{4}}-\frac{99}{28\,y^{3}}\right) +q^{8}\,\left( \frac{218839}{%
17920\,y^{7}}-\frac{1097669}{53760\,y^{6}}+\frac{11479}{1120\,y^{5}}-\frac{%
3891}{2240\,y^{4}}\right)  \nonumber \\
&&\left. -q^{10}\,\left( \frac{318457}{107520\,y^{8}}-\frac{33919}{%
6720\,y^{7}}+\frac{2639}{1024\,y^{6}}-\frac{273}{640\,y^{5}}\right)
+q^{12}\,\left( \frac{24685}{86016\,y^{9}}-\frac{42557}{86016\,y^{8}}+\frac{%
3139}{12288\,y^{7}}-\frac{515}{12288\,y^{6}}\right) \right]  \nonumber \\
&&
\end{eqnarray}

%%Trr
\begin{eqnarray}
\langle T_{r}^{r}\rangle &=&%
%TCIMACRO{\dfrac{p}{\left( 2y-q^{2}\right) ^{6}}}%
%BeginExpansion
{\displaystyle{p \over \left( 2y-q^{2}\right) ^{6}}}%
%EndExpansion
\left[ \frac{21}{10\,y^{2}}-\frac{47}{30\,y^{3}}+q^{2}\,\left( \frac{767}{%
210\,y^{4}}-\frac{1081}{210\,y^{3}}+\frac{9}{35\,y^{2}}\right)
-q^{4}\,\left( \frac{2951}{420\,y^{5}}-\frac{5851}{420\,y^{4}}+\frac{85}{%
12\,y^{3}}-\frac{207}{140\,y^{2}}\right) \right.  \nonumber \\
&&+q^{6}\,\left( \frac{20267}{3360\,y^{6}}-\frac{1469}{112\,y^{5}}+\frac{543%
}{70\,y^{4}}-\frac{207}{140\,y^{3}}\right) -q^{8}\,\left( \frac{26933}{%
10752\,y^{7}}-\frac{101963}{17920\,y^{6}}+\frac{23609}{6720\,y^{5}}-\frac{203%
}{320\,y^{4}}\right)  \nonumber \\
&&\left. +q^{10}\,\left( \frac{53287}{107520\,y^{8}}-\frac{31429}{%
26880\,y^{7}}+\frac{26559}{35840\,y^{6}}-\frac{593}{4480\,y^{5}}\right)
-q^{12}\,\left( \frac{5219}{143360\,y^{9}}-\frac{5531}{61440\,y^{8}}+\frac{%
715}{12288\,y^{7}}-\frac{127}{12288\,y^{6}}\right) \right]  \nonumber \\
&&
\end{eqnarray}

and 
\begin{eqnarray}
&\langle T_{\theta }^{\theta }\rangle =&%
%TCIMACRO{\dfrac{p}{\left( 2y-q^{2}\right) ^{6}}}%
%BeginExpansion
{\displaystyle{p \over \left( 2y-q^{2}\right) ^{6}}}%
%EndExpansion
\left[ \frac{1543}{210\,y^{3}}-\frac{63}{10\,y^{2}}-q^{2}\,\left( \frac{649}{%
35\,y^{4}}-\frac{3509}{210\,y^{3}}+\frac{27}{35\,y^{2}}\right)
+q^{4}\,\left( \frac{40}{y^{5}}-\frac{2049}{35\,y^{4}}+\frac{311}{12\,y^{3}}-%
\frac{621}{140\,y^{2}}\right) \right.  \nonumber \\
&&-q^{6}\,\left( \frac{262747}{6720\,y^{6}}-\frac{210673}{3360\,y^{5}}+\frac{%
1049}{35\,y^{4}}-\frac{621}{140\,y^{3}}\right) +q^{8}\,\left( \frac{342549}{%
17920\,y^{7}}-\frac{1695223}{53760\,y^{6}}+\frac{68787}{4480\,y^{5}}-\frac{%
607}{280\,y^{4}}\right)  \nonumber \\
&&\left. -q^{10}\,\left( \frac{33375}{7168\,y^{8}}-\frac{69651}{8960\,y^{7}}+%
\frac{410303}{107520\,y^{6}}-\frac{593}{1120\,y^{5}}\right) +q^{12}\,\left( 
\frac{194797}{430080\,y^{9}}-\frac{46747}{61440\,y^{8}}+\frac{1537}{%
4096\,y^{7}}-\frac{635}{12288\,y^{6}}\right) \right]  \nonumber \\
&&
\end{eqnarray}

The qualitative behaviour of the stress-energy tensor of the minimally
coupled scalar field is similar to the conformally coupled case, and, once again, 
for the intermediate values of $q$
one has quantitative similarities with the Reissner-Nordstr\"{o}m case. 
Moreover, from figure 4 one can easily deduce the general 
behaviour of the stress-energy tensor for arbitrary coupling for $q < 0.9 .$

%%%%%%%%%%%% ********* rysunek4 ***************
%%% ********************* figure 4 *****************************************
\begin{figure}[tbh]
\centering
\includegraphics[width=12.5cm]{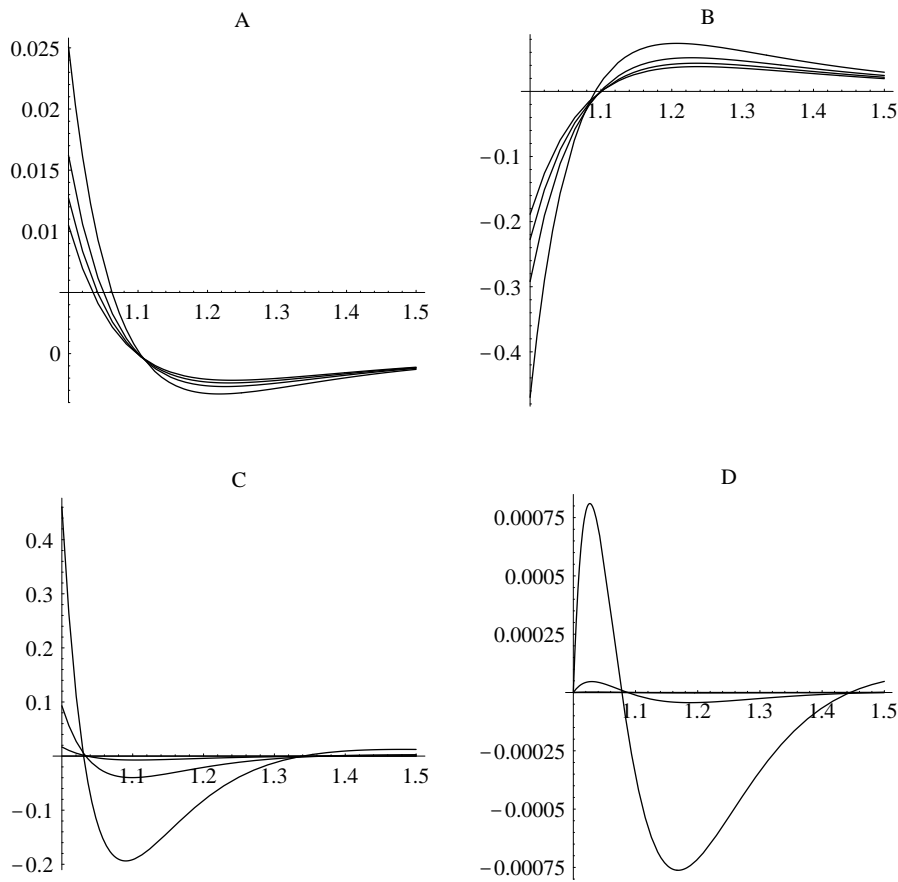}
\caption[aaa]{This graph shows radial dependence of the rescaled $%
T_{t}^{(0)t}$ (panel A), $T_{t}^{(1)t}$ (panel B), $T_{t}^{(2)t}$ (panel C)
and $T_{t}^{(3)t}$ (panel D) for $q\,=\,0.1,\,0.5,\,0.7\,{\rm and}\,0.9.$
The scaling factor is $960\protect\pi ^{2}m^{2}M^{6}.$ The magnitude of $%
T_{t}^{(i)t}$ grows with increasing $q$ for $i\,=\,0,\,2,$ and $3.$ }
\label{fig4}
\end{figure}
%%%%%%%%%%%%%%%%%%%%%%%%%%%%%%%%%%%%%%%%%%%%%%%%%%%%%%%%%%%%%%%%%%%%%%%%%%%%

Finally we remark, that the dilatonic black holes with $a=1$ or $a=0$ do not
exhaust physically important solutions. For example for $a=\sqrt{3}$ one has
a four dimensional effective model reduced from the Kaluza-Klein theory in
five dimensions. By (\ref{schema}) and the approximate stress-energy tensor
expressed in term of $x$, $x_{+}$ and $x_{-}$ could be schematically written
as 
\begin{equation}
\langle T_{a}^{b}\rangle =%
%TCIMACRO{\dfrac{p}{\left[ x\left( x-x_{-}\right) \right] ^{15/2}}}%
%BeginExpansion
{\displaystyle{p \over \left[ x\left( x-x_{-}\right) \right] ^{15/2}}}%
%EndExpansion
\sum_{ijk}d_{ijkb}^{a}\left[ \eta \right] \,x^{i}x_{+}^{j}x_{-}^{k}
\end{equation}
where $0\leq i\leq 7$, $0\leq j\leq 3$ and $0\leq k\leq 6$ subjected to the
condition $i+j+k=9.$ The qualitative behaviour of the stress-energy tensor
for both $\eta =0$ and $\eta =-1/6$ is similar to $\langle T_{a}^{b}\rangle $
constructed in the geometry of a dilatonic black hole with $a=1$ and its run
for small $q$ could be easily inferred form (\ref{ser1}). 

\section{Concluding remarks}

In this paper we have constructed and examined the approximate renormalized
stress-energy tensor of the massive scalar field in the spacetime of the
static electrically charged dilatonic black hole with the special emphasis
put on the string inspired case $a\,=\,1.$ The method employed here is based
on the observation that the lowest order of the expansion of the effective
action in $m^{-2\text{ }}$ could be expressed in terms of the integrated
coincidence limit of coefficient $a_{3}\left( x,x^{\prime }\right) .$
Although the line element of the dilatonic black hole has a simple form, the
analytical formulas describing the stress-energy tensor for a general $a$
constructed within the Schwinger-DeWitt framework are extremely complicated
and hence hard to utilize. Fortunately, for a concrete choice of $a$ there
are considerable simplification.

Expanding for $q\ll 1$ the stress-energy tensor into a power series it is
possible to analyse the influence of $a$ on $\langle T_{a}^{b}\rangle .$ For 
$q=0$ it reduces to the result derived by Frolov and Zel'nikov whereas for
small values of $\tilde{q}$ the stress-energy tensor resembles that
evaluated in the Reissner-Nordstr\"{o}m geometry. The discrepancies between
the tensors grow with $\tilde{q}.$ It should be stressed however that in the
opposite limit the Schwinger-DeWitt technique is inapplicable.

The problem of the massless fields certainly deserves separate treatment,
this however would require extensive numerical calculations as even for
simplest case of the Schwarzschild geometry existing analytical
approximations give, at best, only qualitative agreement with the exact
results. At the moment we only know that the horizon value of the field
fluctuation \cite{kyioshi} 
\begin{equation}
\langle \phi ^{2}\rangle =%
%TCIMACRO{\dfrac{1}{48\pi ^{2}M^{2}x_{+}^{2}}}%
%BeginExpansion
{\displaystyle{1 \over 48\pi ^{2}M^{2}x_{+}^{2}}}%
%EndExpansion
\left[ 1-%
%TCIMACRO{\dfrac{x_{-}}{\left( 1+a^{2}\right) \,x_{+}}}%
%BeginExpansion
{\displaystyle{x_{-} \over \left( 1+a^{2}\right) \,x_{+}}}%
%EndExpansion
\right] \left( 1-%
%TCIMACRO{\dfrac{x_{-}}{x_{+}}}%
%BeginExpansion
{\displaystyle{x_{-} \over x_{+}}}%
%EndExpansion
\right) ^{-%
%TCIMACRO{\tfrac{2a^{2}}{1+a^{2}}}%
%BeginExpansion
{\textstyle{2a^{2} \over 1+a^{2}}}%
%EndExpansion
},
\end{equation}
which is divergent in the extremality limit for $a>0.$ This suggests that
the stress-energy tensor is also divergent at $r_{+}$ of the extremal case.
On the other hand, a first nonvanishing term of the approximation to the
field fluctuation for a massive field is simply 
\begin{equation}
\langle \phi ^{2}\rangle =%
%TCIMACRO{\dfrac{1}{16\pi ^{2}m^{2}}}%
%BeginExpansion
{\displaystyle{1 \over 16\pi ^{2}m^{2}}}%
%EndExpansion
\left[ a_{2}\right] +{\cal O}\left( m^{-4}\right) ,
\end{equation}
and it could be easily shown that 
\begin{equation}
\langle \phi ^{2}\rangle =%
%TCIMACRO{
%\dfrac{f\left( a,r_{+},r_{-}\right) }{720\pi ^{2}m^{2}M^{4}x_{+}^{6}}}%
%BeginExpansion
{\displaystyle{f\left( a,r_{+},r_{-}\right)  \over 720\pi ^{2}m^{2}M^{4}x_{+}^{6}}}%
%EndExpansion
\left( 1+a^{2}\right) ^{-2}\left( 1-%
%TCIMACRO{\dfrac{x_{-}}{x_{+}}}%
%BeginExpansion
{\displaystyle{x_{-} \over x_{+}}}%
%EndExpansion
\right) ^{-%
%TCIMACRO{\tfrac{4a^{2}}{1+a^{2}}}%
%BeginExpansion
{\textstyle{4a^{2} \over 1+a^{2}}}%
%EndExpansion
}+{\cal O}\left( m^{-4}\right) ,
\end{equation}
where 
\begin{equation}
f\left( a,r_{+},r_{-}\right) =\left[ 4+3a^{2}\left( 1-5\xi \right) \right]
x_{-}^{2}-6\left( 1+a^{2}\right) x_{+}x_{-}+3\left( 1+a^{2}\right)
^{2}x_{+}^{2}.
\end{equation}

Finally, we make some comments regarding applications and generalizations of
the results presented in this paper. The question of the massless field has
been addressed above. A careful analysis carried out for $a=0$ in ref.~\cite
{tomimatsu} shows that at least up to ${\cal O}\left( m^{-4}\right) $ the
adapted method approximates well the field fluctuation of the massive field
in the thermal state of temperature $T_{H}.$ It would be interesting to
extend this analysis for any value of $a.$ We also remark that the derived
stress-energy tensors may be employed as a source term of the semiclassical
Einstein field equations. Indeed, preliminary calculations indicate that it
is possible to construct the solution to the linearized semiclassical
Einstein-Maxwell-dilaton equations. We hope that because of their simplicity
presented results will be of use in subsequent applications. We intend to
return to this group of problems elsewhere.

%%%%%%%%%%%%%
%%%%%%%%%%%%%

\centerline{\bf{APPENDIX}}

\subsection{Coincidence limits of the coefficients $a_{2}\left( x,x^{\prime
}\right) ${\protect\small \ }and $a_{3}\left( x,x^{\prime }\right) $}

In this appendix we list coincidence limits of the coefficients $a_{2}\left(
x,x^{\prime }\right) $ and $a_{3}\left( x,x^{\prime }\right) $ for the
scalar field equation (\ref{wave}). With the normalization employed in this
paper the coefficient $[a_{2}]$ reads

\begin{equation}
\lbrack a_{2}]\,=\,-{\frac{1}{6}}\left( \xi -{\frac{1}{5}}\right) \Box R\,+\,%
{\frac{1}{2}}\left( \xi -{\frac{1}{6}}\right) ^{2}R^{2}\,+\,{\frac{1}{180}}%
\left( R_{abcd}R^{abcd}-R_{ab}R^{ab}\right) ,
\end{equation}
whereas $[a_{3}]$ could be written sa 
\begin{equation}
\lbrack a_{3}]\,=\,{\frac{b_{3}}{7!}}\,+\,{\frac{c_{3}}{360}},
\end{equation}
where 
\begin{eqnarray}
b_{3}\, &=&\,{\frac{35}{9}}R^{3}\,+\,17R_{;p}R^{;p}\,-\,R_{qa;p}R^{qa;q}\,-%
\,4R_{qa;p}R^{pa;q}  \nonumber \\
&+&9R_{qabc;p}R^{qabc;p}\,+\,2R\Box R\,+\,18\Box ^{2}R\,-\,8R_{pq}\Box
R^{pq}-{\frac{14}{3}}RR_{pq}R^{pq}  \nonumber \\
&+&24R_{pq;a}^{\ \ \ q}R^{pa}\,-\,{\frac{208}{9}}R_{pq}R^{qa}R_{a}^{\
p}\,+\,12\Box R_{pqab}R^{pqab}+{\frac{64}{3}}R_{pq}R_{ab}R^{paqb}  \nonumber
\\
&-&{\frac{16}{3}}R_{pq}R_{~abc}^{p}R^{qabc}\,+\,{\frac{80}{9}}%
R_{pqab}R_{c~d}^{~p~a}R^{qcbd}\,+\,{\frac{44}{9}}%
R_{pqab}R_{cd}^{~~pq}R^{abcd}
\end{eqnarray}
and 
\begin{eqnarray}
c_{3}\, &=&\,-(5\xi -30\xi ^{2}+60\xi ^{3})R^{3}\,-\,(12\xi -30\xi
^{2})R_{;p}R^{;p}\,-\,(22\xi -60\xi ^{2})R\Box R  \nonumber \\
&-&\,6\xi \Box ^{2}R\,-\,4\xi R_{pq}R^{pq}\,+\,2\xi RR_{pq}R^{pq}\,-\,2\xi
RR_{pqab}R^{pqab}.
\end{eqnarray}

\subsection{$\langle T_{a}^{b}\rangle $ of the massive scalar fields in the
spacetime of the Reissner-Nordstr\"{o}m black hole}

Inserting curvature tensor and its covariant derivatives into the general
formulas obtained form functional differentiation of the effective action (%
\ref{ww}) with respect to the metric tensor one obtains the approximate
stress-energy tensor of massive fields. Since the curvature scalar of the
Reissner-Nordstr\"{o}m geometry is zero, one expects considerable
simplifications. Indeed, it could be easily shown that the tensors ${\tilde{T%
}}_{a}^{(1)b}$ and ${\tilde{T}}_{a}^{(3)b}$ do not contribute to the final
result. The stress-energy tensor of the massive scalar field with arbitrary
coupling with curvature in the Reissner-Nordstr\"{o}m has the form (\ref
{setrn}), where 
\begin{equation}
T_{t}^{(0)t}\,=\,\frac{313\,x^{3}}{7}-\frac{285\,x^{4}}{14}+q^{2}\,\left( 
\frac{-769\,x^{2}}{14}-\frac{192\,x^{3}}{7}+\frac{135\,x^{4}}{7}\right)
+q^{4}\,\left( \frac{514\,x}{7}-\frac{101\,x^{2}}{21}\right) -\frac{%
208\,q^{6}}{7},
\end{equation}
\begin{equation}
T_{r}^{((0)r}\,=-11\,x^{3}+\frac{15\,x^{4}}{2}+q^{2}\,\left( \frac{709\,x^{2}%
}{14}-\frac{248\,x^{3}}{7}+\frac{27\,x^{4}}{7}\right) +q^{4}\,\left( -46\,x+%
\frac{421\,x^{2}}{21}\right) +\frac{74\,q^{6}}{7},
\end{equation}
\begin{equation}
T_{\theta }^{(0)\theta }\,=\,\frac{367\,x^{3}}{7}-\frac{45\,x^{4}}{2}%
+q^{2}\,\left( \frac{-3303\,x^{2}}{14}+\frac{814\,x^{3}}{7}-\frac{81\,x^{4}}{%
7}\right) +q^{4}\,\left( \frac{1726\,x}{7}-\frac{1522\,x^{2}}{21}\right)
-73\,q^{6},
\end{equation}
\begin{equation}
T_{t}^{(1)t}\,=\,-792\,x^{3}+360\,x^{4}+q^{2}\,\left(
2604\,x^{2}-1008\,x^{3}\right) +q^{4}\,\left( -2712\,x+728\,x^{2}\right)
+819\,q^{6},
\end{equation}
\begin{equation}
T_{r}^{(1)r}\,=\,216\,x^{3}-144\,x^{4}+q^{2}\,\left(
-588\,x^{2}+336\,x^{3}\right) +q^{4}\,\left( 504\,x-208\,x^{2}\right)
-117\,q^{6}
\end{equation}
and 
\begin{equation}
T_{\theta }^{(1)\theta }\,=-1008\,x^{3}+432\,x^{4}+q^{2}\,\left(
3276\,x^{2}-1176\,x^{3}\right) +q^{4}\,\left( -3408\,x+832\,x^{2}\right)
+1053\,q^{6}.
\end{equation}

\subsection{Power expansion of the stress-energy tensor for $q\ll 1$}

Repeating the calculations for the line element (\ref{el1}-\ref{el2}) one
obtains components of the stress-energy tensor in the geometry of a general
dilatonic black hole. Assuming $q\ll 1$ and expanding the result into a
power series, after the necessary simplifications (\ref{ser1}), where the
coefficients $t_{a}^{\left( i\right) b}$ are given by 
\begin{equation}
t_{t}^{\left( 1\right) t}=\frac{939}{35}-\frac{76\,x}{7}+\eta \,\left(
192\,x-\frac{2376}{5}\right) ,
\end{equation}
\begin{equation}
t_{r}^{\left( 1\right) r}=4x-\frac{33}{5}+\eta \,\left( \frac{648}{5}-\frac{%
384\,x}{5}\right) ,
\end{equation}
\begin{equation}
t_{\theta }^{\left( 1\right) \theta }=\frac{1101}{35}-12\,x-\eta \,\left( 
\frac{3024}{5}-\frac{1152\,x}{5}\right) ,
\end{equation}
\begin{eqnarray}
t_{t}^{\left( 2\right) t} &=&\frac{1207}{140}-\frac{4359\,a^{2}}{140}-\frac{%
9773}{210\,x}+\frac{939\,a^{2}}{14\,x}+\frac{181\,x}{105}+\frac{19\,a^{2}\,x%
}{7}-\frac{x^{2}}{14}  \nonumber \\
&&+\eta \,\left( \frac{2754\,a^{2}}{5}-\frac{1594}{15}+\frac{888}{x}-\frac{%
1188\,a^{2}}{x}-\frac{436\,x}{5}-48\,a^{2}\,x+12\,x^{2}\right)  \nonumber \\
&&+{\eta }^{2}\,\left( -6076+\frac{6000}{x}+1908\,x-180\,x^{2}\right) , 
\nonumber \\
&&
\end{eqnarray}
\begin{eqnarray}
t_{r}^{\left( 2\right) r} &=&-\frac{12091}{1260}+\frac{213\,a^{2}}{20}+\frac{%
3793}{210\,x}-\frac{33\,a^{2}}{2\,x}+\frac{83\,x}{315}-a^{2}\,x+\frac{%
11\,x^{2}}{70}  \nonumber \\
&&+\eta \,\left( 58-\frac{1026\,a^{2}}{5}-\frac{1032}{5\,x}+\frac{324\,a^{2}%
}{x}+\frac{388\,x}{15}+\frac{96\,a^{2}\,x}{5}-\frac{24\,x^{2}}{5}\right) 
\nonumber \\
&&+{\eta }^{2}\,\left( 1556-\frac{1200}{x}-612\,x+72\,x^{2}\right)  \nonumber
\\
&&
\end{eqnarray}
\begin{eqnarray}
t_{\theta }^{\left( 2\right) \theta } &=&\frac{6841}{252}-\frac{4881\,a^{2}}{%
140}-\frac{18139}{210\,x}+\frac{1101\,a^{2}}{14\,x}+\frac{227\,x}{315}%
+3\,a^{2}\,x-\frac{33\,x^{2}}{70}  \nonumber \\
&&+\eta \,\left( \frac{3348\,a^{2}}{5}+\frac{5572}{5\,x}-\frac{1512\,a^{2}}{x%
}-\frac{1628\,x}{15}-\frac{288\,a^{2}\,x}{5}+\frac{72\,x^{2}}{5}-\frac{296}{3%
}\right)  \nonumber \\
&&+{\eta }^{2}\,\left( -7232+\frac{7200}{x}+2268\,x-216\,x^{2}\right) , 
\nonumber \\
&&
\end{eqnarray}
\begin{eqnarray}
t_{t}^{\left( 3\right) t} &=&\frac{4847}{840}-\frac{4138\,a^{2}}{315}+\frac{%
4359\,a^{4}}{280}+\frac{20149}{420\,x^{2}}-\frac{24971\,a^{2}}{210\,x^{2}}+%
\frac{3443\,a^{4}}{28\,x^{2}}-\frac{9059}{420\,x}+\frac{6026\,a^{2}}{105\,x}-%
\frac{297\,a^{4}}{4\,x}-\frac{79\,x}{210}  \nonumber \\
&&+\frac{152\,a^{2}\,x}{105}-\frac{19\,a^{4}\,x}{14}-\frac{x^{2}}{28}+\frac{%
a^{2}\,x^{2}}{28}+\eta \left( \frac{5642\,a^{2}}{5\,x^{2}}-\frac{1294}{15}-%
\frac{662\,a^{2}}{5}-\frac{1377\,a^{4}}{5}-\frac{4346}{5\,x^{2}}-\frac{%
2178\,a^{4}}{x^{2}}\right.  \nonumber \\
&&\left. +\frac{1936}{5\,x}-\frac{278\,a^{2}}{15\,x}+\frac{1314\,a^{4}}{x}-%
\frac{78\,x}{5}+\frac{198\,a^{2}\,x}{5}+24\,a^{4}\,x+6\,x^{2}-6\,a^{2}%
\,x^{2}\right)  \nonumber \\
&&+\eta ^{2}\left( 7634\,a^{2}-658-\frac{2232}{x^{2}}+\frac{21984\,a^{2}}{%
x^{2}}+\frac{696}{x}-\frac{21596\,a^{2}}{x}+534\,x-1254\,a^{2}\,x-90%
\,x^{2}+90\,a^{2}\,x^{2}\right)  \nonumber \\
&&
\end{eqnarray}
\begin{eqnarray}
t_{r}^{\left( 3\right) r} &=&-\frac{1783}{840}+\frac{187\,a^{2}}{70}-\frac{%
213\,a^{4}}{40}-\frac{5009}{420\,x^{2}}+\frac{5111\,a^{2}}{210\,x^{2}}-\frac{%
121\,a^{4}}{4\,x^{2}}+\frac{401}{60\,x}-\frac{4352\,a^{2}}{315\,x}+\frac{%
93\,a^{4}}{4\,x}  \nonumber \\
&&+\frac{109\,x}{630}-\frac{2\,a^{2}\,x}{45}+\frac{a^{4}\,x}{2}+\frac{%
11\,x^{2}}{140}-\frac{11\,a^{2}\,x^{2}}{140}+\eta \left( \frac{86}{3}+\frac{%
188\,a^{2}}{5}+\frac{513\,a^{4}}{5}+\frac{158}{x^{2}}-\frac{250\,a^{2}}{x^{2}%
}\right.  \nonumber \\
&&\left. +\frac{594\,a^{4}}{x^{2}}-\frac{96}{x}+\frac{118\,a^{2}}{5\,x}-%
\frac{450\,a^{4}}{x}+\frac{18\,x}{5}-\frac{66\,a^{2}\,x}{5}-\frac{%
48\,a^{4}\,x}{5}-\frac{12\,x^{2}}{5}+\frac{12\,a^{2}\,x^{2}}{5}\right) 
\nonumber \\
&&+{\eta }^{2}\,\left( 130-2442\,a^{2}+\frac{288}{x^{2}}-\frac{4656\,a^{2}}{%
x^{2}}-\frac{60}{x}+\frac{5764\,a^{2}}{x}-166\,x+454\,a^{2}\,x+36\,x^{2}-36%
\,a^{2}\,x^{2}\right) .  \nonumber \\
&&
\end{eqnarray}
and 
\begin{eqnarray}
t_{\theta }^{\left( 3\right) \theta } &=&\frac{2411}{360}-\frac{7267\,a^{2}}{%
1260}+\frac{4881\,a^{4}}{280}+\frac{29963}{420\,x^{2}}-\frac{8857\,a^{2}}{%
70\,x^{2}}+\frac{4037\,a^{4}}{28\,x^{2}}-\frac{11617}{420\,x}+\frac{%
2792\,a^{2}}{63\,x}-\frac{2361\,a^{4}}{28\,x}  \nonumber \\
&&-\frac{13\,x}{315}-\frac{31\,a^{2}\,x}{90}-\frac{3\,a^{4}\,x}{2}-\frac{%
33\,x^{2}}{140}+\frac{33\,a^{2}\,x^{2}}{140}+\eta \left( \frac{1378\,a^{2}}{%
x^{2}}-\frac{1252}{15}-\frac{926\,a^{2}}{5}-\frac{1674\,a^{4}}{5}-\frac{5214%
}{5\,x^{2}}\right.  \nonumber \\
&&\left. -\frac{2772\,a^{4}}{x^{2}}+\frac{2052}{5\,x}+\frac{976\,a^{2}}{15\,x%
}+\frac{1620\,a^{4}}{x}-\frac{108\,x}{5}+\frac{252\,a^{2}\,x}{5}+\frac{%
144\,a^{4}\,x}{5}+\frac{36\,x^{2}}{5}-\frac{36\,a^{2}\,x^{2}}{5}\right) 
\nonumber \\
&&+{\eta }^{2}\,\left( 9182\,a^{2}-870-\frac{2970}{x^{2}}+\frac{26640\,a^{2}%
}{x^{2}}+\frac{1098}{x}-\frac{26008\,a^{2}}{x}+644\,x-1508\,a^{2}\,x-108%
\,x^{2}+108\,a^{2}\,x^{2}\right)  \nonumber \\
&&
\end{eqnarray}
The terms containing $\eta ^{3}$ appear starting form $q^{8}$

\end{document}